\documentclass[notitlepage,showkeys,nofootinbib]{revtex4-1}
\usepackage{amsmath,amssymb,bm,graphicx,hyperref}
\allowdisplaybreaks[1]

\renewcommand\d{\partial}
\newcommand\+{\dagger}
\newcommand\<{\langle}
\renewcommand\>{\rangle}
\newcommand\up{\uparrow}
\newcommand\down{\downarrow}
\renewcommand\H{\bm{H}}
\renewcommand\j{\bm{j}}
\renewcommand\k{\bm{k}}
\newcommand\n{\bm{n}}
\newcommand\x{\bm{x}}
\newcommand\X{\bm{X}}
\newcommand\z{\bm{z}}
\renewcommand\L{\mathcal{L}}
\newcommand\R{\mathcal{R}}
\newcommand\eff{\mathrm{eff}}
\newcommand\kF{k_\mathrm{F}}
\newcommand\sgn{\mathrm{sgn}}
\newcommand\tr{\mathrm{tr}}
\newcommand\SO{\mathrm{SO}}
\newcommand\SU{\mathrm{SU}}
\newcommand\U{\mathrm{U}}

\begin{document}

\title{Low-energy effective field theory of superfluid $^3$He-B\\
and its gyromagnetic and Hall responses}

\author{Keisuke Fujii}
\author{Yusuke Nishida}
\affiliation{Department of Physics, Tokyo Institute of Technology,
Ookayama, Meguro, Tokyo 152-8551, Japan}

\date{October 2016}

\begin{abstract}
The low-energy physics of a superfluid $^3$He-B is governed by Nambu--Goldstone bosons resulting from its characteristic symmetry breaking pattern.
Here we construct an effective field theory at zero temperature consistent with all available symmetries in curved space, which are the $\U(1)_\text{phase}\times\SU(2)_\text{spin}\times\SO(3)_\text{orbital}$ gauge invariance and the nonrelativistic general coordinate invariance, up to the next-to-leading order in a derivative expansion.
The obtained low-energy effective field theory is capable of reproducing gyromagnetic responses of the superfluid $^3$He-B, such as a magnetization generated by a rotation and an orbital angular momentum density generated by a magnetic field, in a model-independent and nonperturbative way.
We furthermore show that the stress tensor exhibits a dissipationless Hall viscosity with coefficients uniquely fixed by the orbital angular momentum density, which manifests itself as an elliptical polarization of sound wave with an induced transverse component.
\end{abstract}


\maketitle
\tableofcontents

\section{Introduction}
Since its discovery in 1972, the superfluid $^3$He has been one of the most fascinating systems in condensed matter physics~\cite{Osheroff:1972a,Osheroff:1972b}.
Because the Cooper pair condenses in the spin-triplet and $p$-orbital channel, resulting internal degrees of freedom enrich the phase diagram significantly by making different symmetry breaking patterns possible~\cite{Vollhardt:1990}.
Our understanding developed in the superfluid $^3$He has contributed extensively to other unconventional superconductivity not only in condensed matter systems~\cite{Sigrist:1991,Mackenzie:2003} but also in dense nuclear matter~\cite{Lombardo:1999,Alford:2008}.
Furthermore, various parallels may make the superfluid $^3$He serve as an analog simulator of high-energy phenomena~\cite{Volovik:2003}.

In spite of such long-standing study, our fascination for the superfluid $^3$He never ends.
Recently, it has received renewed interest from the perspective of topology~\cite{Mizushima:2015,Mizushima:2016}.
A thin film of the Anderson--Brinkman--Morel state ($^3$He-A) is a prototype of topological superfluids in two dimensions belonging to class D, while the Balian--Werthamer state ($^3$He-B) is a prototype of topological superfluids in three dimensions belonging to class DIII~\cite{Schnyder:2008}.
Emergent surface Andreev bound states in the latter have been probed experimentally, which can be regarded as the celebrated Majorana fermions~\cite{Wilczek:2009,Okuda:2012}.

While many fascinating features are exhibited by the superfluid $^3$He, its quantitatively reliable analysis is often challenging because the system is strongly correlated.
One powerful approach suitable for such strongly correlated superfluids is the effective field theory~\cite{Weinberg:1996}.
When fermions are fully gapped, the low-energy physics of a superfluid is governed by Nambu--Goldstone bosons resulting from spontaneously broken symmetries.
Their effective action consists of an infinite number of terms allowed by the symmetries of the system but organized systematically according to a derivative expansion.

In order to exploit all available symmetry constraints, it is advantageous to couple the system with external gauge fields to temporarily promote the global symmetries to their local counterparts.
In particular, the Galilean invariance is promoted to the nonrelativistic general coordinate invariance by putting the system in curved space with an external metric~\cite{Son:2006}.
Such an approach proves to be powerful to obtain nontrivial outcomes even in flat space as demonstrated previously for the unitary Fermi gas~\cite{Son:2006,Son:2007} and for the chiral superfluid in two dimensions~\cite{Hoyos:2014a}.
In this paper, we employ the same approach to shed new light on intriguing physics of the superfluid $^3$He-B, where a rotation and a magnetic field generate a magnetization and an orbital angular momentum density, respectively, together with a novel Hall viscosity.
While the gyromagnetic responses can be understood intuitively by the spin--orbit locking in the superfluid $^3$He-B~\cite{Salomaa:1987}, our predictions resulting from the symmetries alone are model-independent, nonperturbative, and hence quantitatively reliable.

To this end, we first review the symmetries and their spontaneous breaking in the superfluid $^3$He-B and discuss how they can be promoted to the local symmetries in curved space in Sec.~\ref{sec:symmetry}.
We then construct an effective field theory at zero temperature up to the next-to-leading order in a derivative expansion in Sec.~\ref{sec:effective}.
Its physical consequences are investigated in Sec.~\ref{sec:response} and, in particular, we predict the gyromagnetic and Hall responses of the superfluid $^3$He-B.
Finally, we summarize and conclude this paper in Sec.~\ref{sec:summary} and some detailed information is presented in Appendix.

In what follows, we shall set $\hbar=\mu_0=1$ with $\mu_0$ being the vacuum permeability and employ shorthand notations $(x)=(t,\x)=(t,x^1,x^2,x^3)$ for spacetime coordinates and $\phi\,\tensor\d_{\!\mu}\psi\equiv[\phi(\d_\mu\psi)-(\d_\mu\phi)\psi]/2$ with $\mu=t,1,2,3$.
In and after Sec.~\ref{sec:curved}, $\alpha,\beta,\gamma,\dots$ and $a,b,c,\dots$ refer to spin and orbital indices, respectively, with no distinction between upper and lower indices for them.
In contrast, $i,j,k,\dots$ refer to general coordinate indices in curved space, which are lowered and raised by a metric $g_{ij}(x)$ and its inverse $g^{ij}(x)$.
We partly follow standard notations in general relativity such as for the metric determinant denoted by $g(x)\equiv\det[g_{ij}(x)]$, the Christoffel symbols by $\Gamma^i_{jk}(x)\equiv g^{il}(x)[\d_jg_{kl}(x)+\d_kg_{jl}(x)-\d_lg_{jk}(x)]/2$, the covariant derivatives by $\nabla_iv_j(x)\equiv\d_iv_j(x)-\Gamma_{ij}^k(x)v_k(x)$, $\nabla_iv^j(x)\equiv\d_iv^j(x)+\Gamma_{ik}^j(x)v^k(x)$, and so on~\cite{Nakahara:2003}.
Last but not least, implicit sums of repeated indices over 1, 2, 3 are assumed throughout this paper.

\section{From flat to curved space
\label{sec:symmetry}}
\subsection{Superfluid $^3$He-B in flat space}
The microscopic action describing a liquid $^3$He with slight idealization of neglecting the dipolar coupling is
\begin{align}
S_\text{flat} &= \int\!d^4x\biggl[\psi^\+(x)i\tensor\d_t\psi(x)
- \frac1{2m}\d_i\psi^\+(x)\d_i\psi(x)\biggr] \notag\\
&\quad - \frac12\int\!d^4xd^4x'\psi_\sigma^\+(x)\psi_{\sigma'}^\+(x')
\delta(t-t')V(|\x-\x'|)\psi_{\sigma'}(x')\psi_\sigma(x),
\end{align}
where $\psi(x)=[\psi_\up(x),\psi_\down(x)]^T$ is a spin-1/2 fermion field with its mass $m$ and $V(r)$ is an interatomic potential.
Evidently, it is invariant under the U(1) phase and SU(2) spin rotations, $\psi(x)\to e^{i\phi}S\psi(x)$, as well as the SO(3) spatial rotation $\d_i\psi(x)\to R_{ij}\d_j\psi(x)$.
Therefore, the symmetry group of the microscopic action includes
\begin{align}
G = \U(1)_\phi \times \SU(2)_S \times \SO(3)_L
\end{align}
in addition to the spacetime translation, Galilean boost, parity inversion, and time reversal.

The ground state of the liquid $^3$He is a superfluid state, where the fermion bilinear
\begin{align}
\Delta_i(x) = \<\psi^T(x)\sigma_2\sigma_\alpha\tensor\d_i\psi(x)\>\sigma_\alpha
\end{align}
in the spin-triplet ($\alpha=1,2,3$) and $p$-orbital ($i=1,2,3$) channel acquires a nonzero vacuum expectation value.
Here the spin index of the fermion bilinear is contracted by multiplying the outer Pauli matrix to make the order parameter matrix-valued and transform as $\Delta_i(x)\to e^{2i\phi}SR_{ij}\Delta_j(x)S^\+$.
In particular, the order parameter of the superfluid $^3$He-B is provided by a constant $\Delta_i(x)=\Delta_0\sigma_i$, which transforms as $\Delta_0\sigma_i\to\Delta_0e^{2i\phi}SL^\+\sigma_iLS^\+$ with $L$ being the SU(2) matrix related to the rotation matrix by $L^\+\sigma_iL=R_{ij}\sigma_j$.
Therefore, the spin and orbital degrees of freedom are locked by the pair condensation so that the symmetry group $G$ is spontaneously broken down to SU(2)$_{S+L}$.

Because such spontaneously broken symmetry transformation does not change the energy of the system, the ground state is continuously degenerate.
The spacetime variation of the order parameter in the ground-state manifold thus describes gapless excitations, which is parametrized by
\begin{align}
\Delta_i(x) = \Delta_0e^{2i\theta(x)}U(x)\sigma_iU^\+(x).
\end{align}
Here $\theta(x)$ and $U(x)\in\SU(2)$ are Nambu--Goldstone fields of the superfluid $^3$He-B corresponding to the phase and spin--orbit collective modes, respectively, and transform as $\theta(x)\to\theta(x)+\phi$ and $U(x)\to SU(x)L^\+$.
While the spin--orbit collective mode is usually represented by the rotation matrix~\cite{Vollhardt:1990}, we find the SU(2) representation rather advantageous in later constructing the low-energy effective field theory because of its close parallels to the chiral perturbation theory~\cite{Scherer:2003}.

\subsection{Superfluid $^3$He-B in curved space
\label{sec:curved}}
We now wish to promote the above global symmetries to their local counterparts.
The local U(1)$_\phi$ and SU(2)$_S$ rotations act on the fermion field as
\begin{align}
\psi(x) \to e^{i\phi(x)}S(x)\psi(x).
\end{align}
The microscopic action can be made invariant by coupling it with external U(1) and SU(2) gauge fields through the gauge covariant derivative $D_\mu\psi(x)\equiv[\d_\mu-iA_\mu(x)-iB_\mu(x)]\psi(x)$ and imposing their gauge transformations of
\begin{align}
A_\mu(x) &\to A_\mu(x) + \d_\mu\phi(x), \label{eq:U(1)}\\
B_\mu(x) &\to S(x)B_\mu(x)S^\+(x) - i\d_\mu S(x)S^\+(x), \label{eq:SU(2)}
\end{align}
so that $D_\mu\psi(x)\to e^{i\phi(x)}S(x)D_\mu\psi(x)$.

On the other hand, in order to achieve the local SO(3)$_L$ invariance, we need to put the system in curved space with an external metric $g_{ij}(x)\equiv e_i^a(x)e_j^b(x)\delta_{ab}$.
Here a set of vector fields $e_i^a(x)$ is called the vielbein ($a=1,2,3$) and defines a local orthonormal frame, where the local SO(3)$_L$ rotation acts as
\begin{align}
e_i^a(x) \to R_{ab}(x)e_i^b(x) \label{eq:SO(3)}
\end{align}
and hence $D_a\psi(x)\equiv e_a^i(x)D_i\psi(x)\to R_{ab}(x)D_b\psi(x)$ for the spatial covariant derivative~\cite{Nakahara:2003}.
Accordingly, the microscopic action in curved space becomes
\begin{align}\label{eq:action}
S_\text{curved} = \int\!d^4x\sqrt{g(x)}
\biggl[\psi^\+(x)[i\tensor\d_t+A_t(x)+B_t(x)]\psi(x)
- \frac1{2m}D_a\psi^\+(x)D_a\psi(x)\biggr] + S_\text{int},
\end{align}
which, including the interaction term as shown later, enjoys the $\U(1)_\phi\times\SU(2)_S\times\SO(3)_L$ gauge invariance.

Similarly, the order parameter of the superfluid $^3$He-B in curved space becomes
\begin{align}
\Delta_a(x) = \<\psi^T(x)\sigma_2\sigma_\alpha\tensor D_a\psi(x)\>\sigma_\alpha,
\end{align}
which in the ground-state manifold is parametrized by the Nambu--Goldstone fields as
\begin{align}
\Delta_a(x) = \Delta_0e^{2i\theta(x)}U(x)\sigma_aU^\+(x).
\end{align}
They transform under the gauge transformation as
\begin{align}
\theta(x) &\to \theta(x) + \phi(x), \\
U(x) &\to S(x)U(x)L^\+(x),
\end{align}
with the same relationship $L^\+(x)\sigma_aL(x)=R_{ab}(x)\sigma_b$ as before.

While the external gauge fields are introduced to grant the gauge invariance to the microscopic action, their physical meanings can be extracted from Eq.~(\ref{eq:action}).
The temporal component of the U(1)$_\phi$ gauge field is equivalent to the chemical potential including a spacetime-dependent trapping potential and that of the SU(2)$_S$ gauge field to the Zeeman energy with a spacetime-dependent magnetic field, both of which can be realized in the superfluid $^3$He-B.
The spatial component of the U(1)$_\phi$ gauge field can also be produced in a rotating frame of reference~\cite{Cooper:2008}, while that of the SU(2)$_S$ gauge field bears no simple realization to our knowledge.

\subsection{Nonrelativistic diffeomorphism}
The microscopic action in curved space furthermore enjoys the nonrelativistic general coordinate invariance.
Under an infinitesimal but spacetime-dependent shift of spatial coordinates $x^i\to x^i+\xi^i(x)$, the microscopic action remains invariant if the fields therein are transformed according to
\begin{align}
\delta\psi(x) &= -\xi^j(x)\d_j\psi(x), \\
\delta A_t(x) &= -\xi^j(x)\d_jA_t(x) - A_j(x)\dot\xi^j(x), \label{eq:NGC_A}\\
\delta A_i(x) &= -\xi^j(x)\d_jA_i(x) - A_j(x)\d_i\xi^j(x) - mg_{ij}(x)\dot\xi^j(x), \\
\delta B_t(x) &= -\xi^j(x)\d_jB_t(x) - B_j(x)\dot\xi^j(x), \\
\delta B_i(x) &= -\xi^j(x)\d_jB_i(x) - B_j(x)\d_i\xi^j(x), \\
\delta e_i^a(x) &= -\xi^j(x)\d_je_i^a(x) - e_j^a(x)\d_i\xi^j(x), \label{eq:NGC_e}
\end{align}
with $\dot\xi(x)\equiv\d_t\xi(x)$~\cite{Son:2006}.
This is called the nonrelativistic general coordinate transformation, which is a local version of the spatial translation including the Galilean boost as a special case.
Fields that transform as $\delta s(x)=-\xi^j(x)\d_js(x)$ and $\delta v_i(x)=-\xi^j(x)\d_jv_i(x)-v_j(x)\d_i\xi^j(x)$ are referred to as scalars and vectors, respectively.
Because the order parameter of the superfluid $^3$He-B in curved space is merely a set of scalar fields, the Nambu--Goldstone fields transform likewise as
\begin{align}
\delta\theta(x) &= -\xi^j(x)\d_j\theta(x), \\
\delta U(x) &= -\xi^j(x)\d_jU(x).
\end{align}

So far we have not specified the interaction term in curved space.
It is required not only to be invariant under the gauge transformation and the general coordinate transformation but also to produce the interatomic potential of $^3$He atoms in flat space, which we shall model by the Lennard-Jones potential $V(r)=C_{12}/r^{12}+C_6/r^6$.
Each power-law potential can be realized by coupling the fermion field with an auxiliary massless field $\varphi(x,\X)$ living in $n+2$ spatial dimensions as
\begin{align}
S_n = \sgn(C_n)\int\!d^4x\sqrt{g(x)}\int\!d^{n-1}\X
\biggl[\frac{g^{ij}(x)}{2}\d_i\varphi(x,\X)\d_j\varphi(x,\X)
+ \frac{\delta^{IJ}}{2}\d_I\varphi(x,\X)\d_J\varphi(x,\X) \notag\\
- \lambda_n\varphi(x,\X)\psi_\sigma^\+(x)\psi_\sigma(x)\delta^{n-1}(\X)\biggr],
\end{align}
where $X^I$ ($I=4,5,\dots,n+2$) are coordinates in extra dimensions~\cite{Hoyos:2012}.
It proves to comply with the required invariance if the auxiliary field is transformed as a scalar.
Also, by integrating out the auxiliary field with the help of its equation of motion in flat space, the power-law potential $V_n(r)=C_n/r^n$ is obtained for a coupling constant $\lambda_n^2=4\pi^{1+n/2}|C_n|/\Gamma(n/2)$.
Therefore, the Lennard-Jones potential in flat space can be incorporated in the microscopic action by the interaction term of $S_\text{int}=S_{12}+S_6$ in a way compatible with the gauge invariance and the general coordinate invariance in curved space.

\section{Low-energy effective action
\label{sec:effective}}
\subsection{Preliminaries}
Because fermions are fully gapped in the superfluid $^3$He-B, its low-energy physics is governed by the Nambu--Goldstone bosons.
By integrating out the fermion and auxiliary fields in curved space, we obtain the effective action written in terms of the Nambu--Goldstone fields, $\theta(x)$ and $U(x)$, as well as the external fields, $A_\mu(x)$, $B_\mu(x)$, and $e_i^a(x)$.
Crucially, it must be consistent with the gauge invariance and the general coordinate invariance of the microscopic action, which is our guiding principle in constructing the low-energy effective field theory.

In order to proceed, we need to introduce a spin connection according to
\begin{align}
\omega_t(x) &\equiv \frac{i}{8}[\sigma^j(x),\d_t\sigma_j(x)]
- \frac{i}{8m}[\sigma^j(x),\sigma^k(x)]\d_jA_k(x), \label{eq:omega_def}\\
\omega_i(x) &\equiv \frac{i}{8}[\sigma^j(x),\nabla_i\sigma_j(x)],
\end{align}
where $\sigma_i(x)\equiv e_i^a(x)\sigma_a$ is a set of Pauli matrices in curved space to obey
\begin{align}
\sigma_i(x)\sigma_j(x) = g_{ij}(x) + i\varepsilon_{ijk}(x)\sigma^k(x)
\end{align}
with $\varepsilon_{ijk}(x)\equiv e_i^a(x)e_j^b(x)e_k^c(x)\epsilon_{abc}=\sqrt{g(x)}\epsilon_{ijk}$ being the antisymmetric tensor~\cite{Nakahara:2003}.
Because $\sigma_i(x)$ transforms as $\sigma_i(x)\to L(x)\sigma_i(x)L^\+(x)$ under the gauge transformation and as a vector under the general coordinate transformation, the spin connection proves to transform as
\begin{align}
\omega_\mu(x) \to L(x)\omega_\mu(x)L^\+(x) - i\d_\mu L(x)L^\+(x)
\end{align}
and as
\begin{align}
\delta\omega_t(x) &= -\xi^j(x)\d_j\omega_t(x) - \omega_j(x)\dot\xi^j(x), \label{eq:omega_trans}\\
\delta\omega_i(x) &= -\xi^j(x)\d_j\omega_i(x) - \omega_j(x)\d_i\xi^j(x),
\end{align}
so that it plays the role of SU(2)$_L$ gauge field.
What is crucial to our later predictions is that the second term in Eq.~(\ref{eq:omega_def}), which is not constrained by the Galilean invariance alone, is uniquely fixed to make the temporal component of the spin connection transform canonically as in Eq.~(\ref{eq:omega_trans}) under the general coordinate transformation~\cite{Hoyos:2014a}.

The gauge fields can appear in the gauge invariant effective action either through gauge covariant derivatives of the Nambu--Goldstone fields,
\begin{align}
D_\mu\theta(x) &\equiv \d_\mu\theta(x) - A_\mu(x), \\
D_\mu U(x) &\equiv \d_\mu U(x) - iB_\mu(x)U(x) + U(x)i\omega_\mu(x),
\end{align}
or through their field strength tensors,
\begin{align}
F_{\mu\nu}(x) &\equiv \d_\mu A_\nu(x) - \d_\nu A_\mu(x), \\
G_{\mu\nu}(x) &\equiv \d_\mu B_\nu(x) - \d_\nu B_\mu(x) - i[B_\mu(x),B_\nu(x)], \\
H_{\mu\nu}(x) &\equiv \d_\mu\omega_\nu(x) - \d_\nu\omega_\mu(x) - i[\omega_\mu(x),\omega_\nu(x)].
\end{align}
Here $D_\mu\theta(x)$ and $F_{\mu\nu}(x)$ are gauge invariant, while the others are gauge covariant and transform as $D_\mu U(x)\to S(x)D_\mu U(x)L^\+(x)$, $G_{\mu\nu}(x)\to S(x)G_{\mu\nu}(x)S^\+(x)$, and $H_{\mu\nu}(x)\to L(x)H_{\mu\nu}(x)L^\+(x)$.

On the other hand, under the general coordinate transformation, the gauge covariant derivatives transform as
\begin{align}
\delta D_t\theta(x) &= -\xi^j(x)\d_jD_t\theta(x) - D_j\theta(x)\dot\xi^j(x), \\
\delta D_i\theta(x) &= -\xi^j(x)\d_jD_i\theta(x) - D_j\theta(x)\d_i\xi^j(x) + mg_{ij}(x)\dot\xi^j(x), \\
\delta D_tU(x) &= -\xi^j(x)\d_jD_tU(x) - D_jU(x)\dot\xi^j(x), \\
\delta D_iU(x) &= -\xi^j(x)\d_jD_iU(x) - D_jU(x)\d_i\xi^j(x).
\end{align}
While $D_iU(x)$ is a vector, $D_t\theta(x)$ and $D_tU(x)$ are not scalars because of the last terms in their transformation laws.
Such undesirable terms can be eliminated with the help of $D_i\theta(x)$ by modifying the temporal covariant derivatives as
\begin{align}
\tilde D_t\theta(x) &\equiv D_t\theta(x) + \frac{g^{ij}(x)}{2m}D_i\theta(x)D_j\theta(x), \\
\tilde D_tU(x) &\equiv D_tU(x) + \frac{g^{ij}(x)}{m}D_i\theta(x)D_jU(x),
\end{align}
which now transform as genuine scalars and are actually equivalent to the material derivatives in fluid mechanics.

\subsection{Power counting scheme}
We are ready to construct the effective action with the building blocks introduced above.
While there is an infinite number of terms allowed by the gauge invariance and the general coordinate invariance, they can be organized systematically according to a derivative expansion valid in the low-energy limit.
Our power counting scheme is such that
\begin{align}
\d_\mu\theta(x),\quad U(x),\quad A_\mu(x),\quad e_i^a(x)
\end{align}
are regarded as $O(1)$, while
\begin{align}
\d_\mu,\quad B_\mu(x)
\end{align}
are regarded as small expansion parameters of $O(p)$.

It is indeed possible and more comprehensive to treat $\d_\mu\theta(x)$ as $O(1)$ in spite of $\d_\mu\sim O(p)$ because $\theta(x)$ does not appear in the effective action without derivatives acting on it~\cite{Son:2005,Son:2006}.
We also assume $A_\mu(x)\sim\d_\mu\theta(x)$ but $B_\mu(x)$ as small as $\d_\mu$ so that the gauge covariant derivatives bear definite power countings, which later means the Zeeman energy being small compared to the chemical potential.
Accordingly, the gauge covariant derivatives and the field strength tensors are granted the power countings of
\begin{align}
D_\mu\theta(x) \quad&\sim\quad O(p^0), \\
D_\mu U(x),\quad F_{\mu\nu}(x) \quad&\sim\quad O(p^1), \\
G_{\mu\nu}(x),\quad H_{\mu\nu}(x) \quad&\sim\quad O(p^2).
\end{align}
At each order in the derivative expansion, only a finite number of terms is allowed in the effective action.
Here, in addition to the gauge invariance and the general coordinate invariance, the parity and time-reversal invariance of the microscopic action must be imposed as well on the effective action of the superfluid $^3$He-B.

\subsection{Construction and calibration}
The effective action is a spacetime integral of the corresponding Lagrangian density,
\begin{align}
S_\eff[\theta,U,A_\mu,B_\mu,e_i^a] = \int\!d^4x\sqrt{g(x)}\,\L_\eff(x),
\end{align}
which can be divided into a part involving only the phase collective mode and the rest involving the spin--orbit collective mode as well.
The former part is actually identical to the effective action of spin-singlet $s$-orbital superfluids, which was already constructed up to the next-to-leading order in Ref.~\cite{Son:2006}.
In particular, the leading-order Lagrangian density is $O(1)$ and is expressed as
\begin{align}
\L_\theta^{(0)}(x) = f_0[-\tilde D_t\theta(x)]
\end{align}
with $f_0[\,*\,]$ being an arbitrary function.
Because no terms at $O(p)$ comply with the required invariance, the next-to-leading order is $O(p^2)$ where four distinct Lagrangian densities contribute to the effective action~\cite{Son:2006}.
They are denoted by $\L_{\theta,1-4}^{(2)}(x)$ whose explicit expressions are provided by Eqs.~(\ref{eq:NLO_theta1})--(\ref{eq:NLO_theta4}) in Appendix.

Similarly, the Lagrangian densities involving the spin--orbit collective mode do not contribute to the effective action at $O(p)$, while five distinct contributions are found at $O(p^2)$.
In particular, only one of them involves the temporal covariant derivative of $U(x)$ and is expressed as
\begin{align}\label{eq:NLO_U0}
\L_{U,0}^{(2)}(x) = \frac12g_0[-\tilde D_t\theta(x)]\tr[\tilde D_tU^\+(x)\tilde D_tU(x)]
\end{align}
with $g_0[\,*\,]$ being an arbitrary function.
The other four Lagrangian densities denoted by $\L_{U,1-4}^{(2)}(x)$, in contrast, involve the spatial covariant derivative of $U(x)$ and their explicit expressions are provided by Eqs.~(\ref{eq:NLO_U1})--(\ref{eq:NLO_U4}) in Appendix.
Accordingly, the most general Lagrangian density up to the next-to-leading order in the derivative expansion proves to be
\begin{align}
\L_\eff(x) = \L_\theta^{(0)}(x) + \sum_{n=1}^4\L_{\theta,n}^{(2)}(x)
+ \sum_{n=0}^4\L_{U,n}^{(2)}(x) + O(p^4),
\end{align}
where next-to-next-to-leading-order corrections are as small as $O(p^4)$.%
\footnote{We note that
\begin{align*}
\L_\theta^{\prime(1)}(x) \sim \d_t\ln\!\sqrt{g(x)} + \frac{g^{ij}(x)}{m}\nabla_iD_j\theta(x)
\underset{\text{EoM}}{\sim} \d_t\tilde D_t\theta(x) + \frac{g^{ij}(x)}{m}D_i\theta(x)\d_j\tilde D_t\theta(x)
\end{align*}
and
\begin{align*}
\L_U^{\prime(1)}(x) \sim \frac{g^{ij}(x)}{2}\tr[\sigma_i(x)U^\+(x)iD_jU(x)]
\end{align*}
multiplied by arbitrary functions of $\tilde D_t\theta(x)$ are $O(p)$ consistent with the gauge invariance and the general coordinate invariance but are incompatible with the time-reversal invariance and the parity invariance, respectively.
In general, by inspecting transformation laws of each building block, one can show that Lagrangian densities bearing odd powers of $p$ are odd either in parity inversion or in time reversal and thus do not appear in our effective action.}

The obtained effective action of the superfluid $^3$He-B depends on the ten unknown functions of $f_{0-4}[\,*\,]$ and $g_{0-4}[\,*\,]$.
While they cannot be constrained further from our perspective of low-energy effective field theory, two of them can be related to common thermodynamic quantities.
To this end, we temporarily consider the uniform ground state in flat space where the Nambu--Goldstone fields are constant and the spatial components of the external gauge fields are turned off.
Their temporal components set to be constant are equivalent to the chemical potential $A_t=\mu$ and the Zeeman energy $B_t=(\gamma/2)\bm\sigma\cdot\H$ as already discussed.
Therefore, the Lagrangian density is reduced to
\begin{align}
\L_\eff = f_0[\mu] + \frac{\gamma^2}{4}g_0[\mu]H^2 + O(H^4),
\end{align}
where $\H$ is an applied magnetic field and $\gamma\approx-2.04\times10^8$/(s\,T) is the gyromagnetic ratio of $^3$He nucleus~\cite{Wheatley:1975}.
Because the Lagrangian density differentiated with respect to $\mu$ and $\H$ produces the particle number density and the magnetization, respectively, it must be identical to the pressure of the superfluid $^3$He-B at zero temperature~\cite{Son:2006}.
In particular, $P_0[\mu]\equiv f_0[\mu]$ is the pressure and $\chi_0[\mu]\equiv\gamma^2g_0[\mu]/2$ is the magnetic susceptibility as functions of the chemical potential at zero magnetic field.
The other functions, however, remain unidentified and thus we shall carefully extract universal predictions that are independent of them.
By construction, our effective action and hence predictions are valid at a sufficiently low-energy and long-wavelength regime compared to the superfluid gap energy and coherence length, respectively.

\section{Gyromagnetic and Hall responses
\label{sec:response}}
\subsection{Continuity equations}
We arrive at the stage of investigating physical consequences of our low-energy effective field theory.
When the effective action is invariant under an infinitesimal transformation of the fields therein by $\delta\theta(x)$, $\delta U(x)$, $\delta A_\mu(x)$, $\delta B_\mu(x)$, and $\delta e_i^a(x)$, we obtain the identity:
\begin{align}
\int\!d^4x\biggl[\frac{\delta S_\eff}{\delta A_\mu(x)}\delta A_\mu(x)
+ \frac{\delta S_\eff}{\delta B_\mu^\alpha(x)}\delta B_\mu^\alpha(x)
+ \frac{\delta S_\eff}{\delta e_i^a(x)}\delta e_i^a(x)\biggr] = 0,
\end{align}
where $\delta S_\eff/\delta\theta(x)$ and $\delta S_\eff/\delta U(x)$ are eliminated with the help of the equations of motion for the Nambu--Goldstone fields and the SU(2)$_S$ gauge field is decomposed as $B_\mu(x)\equiv B_\mu^\alpha(x)\sigma_\alpha$.
This identity applied to the gauge transformation and the general coordinate transformation immediately leads to the continuity equations in curved space.

First of all, the U(1)$_\phi$ gauge invariance with Eq.~(\ref{eq:U(1)}) leads to the mass continuity equation:
\begin{align}\label{eq:mass}
\nabla_t\rho(x) + \nabla_ij^i(x) = 0,
\end{align}
where $\nabla_t\rho(x)\equiv\d_t[\sqrt{g(x)}\rho(x)]/\sqrt{g(x)}$ is introduced and
\begin{align}
\rho(x) &\equiv \frac{m}{\sqrt{g(x)}}\frac{\delta S_\eff}{\delta A_t(x)}, \\
j^i(x) &\equiv \frac{m}{\sqrt{g(x)}}\frac{\delta S_\eff}{\delta A_i(x)}
\end{align}
are the mass density and its flux.
Similarly, the SU(2)$_S$ gauge invariance with Eq.~(\ref{eq:SU(2)}) leads to the spin continuity equation:
\begin{align}
\nabla_ts_\alpha(x) + \nabla_ij_\alpha^i(x)
= 2\epsilon_{\alpha\beta\gamma}s_\beta(x)B_t^\gamma(x)
+ 2\epsilon_{\alpha\beta\gamma}j_\beta^i(x)B_i^\gamma(x),
\end{align}
where
\begin{align}
s_\alpha(x) &\equiv \frac1{2\sqrt{g(x)}}\frac{\delta S_\eff}{\delta B_t^\alpha(x)}, \\
j_\alpha^i(x) &\equiv \frac1{2\sqrt{g(x)}}\frac{\delta S_\eff}{\delta B_i^\alpha(x)}
\end{align}
are the spin density and its flux.

On the other hand, the SO(3)$_L$ gauge invariance with Eq.~(\ref{eq:SO(3)}) leads to the constraint, $\varepsilon_{ijk}(x)T^{jk}(x)=0$, imposing the vanishing antisymmetric part on the stress tensor defined by
\begin{align}
T^{ij}(x) \equiv \frac{e_a^i(x)}{\sqrt{g(x)}}\frac{\delta S_\eff}{\delta e_j^a(x)}.
\end{align}
Finally, the general coordinate invariance with Eqs.~(\ref{eq:NGC_A})--(\ref{eq:NGC_e}) leads to the momentum continuity equation:
\begin{align}\label{eq:momentum}
\nabla_tj_i(x) + \nabla_jT_i^j(x)
= \frac1mF_{it}(x)\rho(x) + \frac1mF_{ij}(x)j^j(x)
+ 2G_{it}^\alpha(x)s_\alpha(x) + 2G_{ij}^\alpha(x)j_\alpha^j(x),
\end{align}
where the momentum density proves to be identical to the mass flux and the SU(2)$_S$ field strength tensor is decomposed as $G_{\mu\nu}(x)\equiv G_{\mu\nu}^\alpha(x)\sigma_\alpha$.

\subsection{Current and orbital angular momentum}
While the continuity equations themselves are rather common, the form of the momentum density therein reflects intriguing physics of the superfluid $^3$He-B.
In terms of the mass and spin densities introduced above as well as the mass and spin superfluid velocities defined by $v^i(x)\equiv[g^{ij}(x)/m]D_j\theta(x)$ and $v_\alpha^i(x)\equiv-[g^{ij}(x)/(2m)]\tr[\sigma_\alpha iD_jU(x)U^\+(x)]$, the momentum density is expressed as
\begin{align}\label{eq:current}
j^i(x) = \rho(x)v^i(x) + 2ms_\alpha(x)v_\alpha^i(x)
+ \frac12\varepsilon^{ijk}(x)\d_j\ell_k(x) + O(f_{2-3}) + O(p^4),
\end{align}
where $O(f_{2-3})$ indicates contributions depending on the unknown functions with their explicit expressions being suppressed.
Of particular interest is the term including
\begin{align}\label{eq:locking}
\ell_i(x) = -e_i^a(x)\R_{a\alpha}^T(x)s_\alpha(x),
\end{align}
which can be identified as an orbital angular momentum density because it, assumed to vanish at infinity, contributes to the orbital angular momentum in flat space as $\int\!d\x\,\epsilon_{ijk}x^jj^k(x)=\int\!d\x\,\ell_i(x)+\cdots$.%
\footnote{Because the stress tensor is symmetric, the orbital angular momentum thus expressed with Cartesian coordinates is conserved automatically when the momentum is conserved.}
Therefore, the circulating mass flow is generated by the spatial variation of the orbital angular momentum density, which is an analog of the Mermin--Muzikar current in the superfluid $^3$He-A~\cite{Cross:1975,Ishikawa:1980,Mermin:1980}.
While the magnitude of the orbital angular momentum density contributing to the momentum density had been controversial between the calculations in Refs.~\cite{Combescot:1980,Dombre:1985,Yip:1986} versus Refs.~\cite{Volovik:1983,Volovik:1984,Mineev:1984,Muzikar:1984,Mineev:1986} and remains unresolved experimentally~\cite{Hakonen:1987,Korhonen:1990,Hakonen:1992}, our expression in Eq.~(\ref{eq:current}) reliable in the long-wavelength limit proves to be consistent with the result of Refs.~\cite{Combescot:1980,Dombre:1985,Yip:1986}.

The rotation matrix $\R_{\alpha\beta}(x)$ appearing in Eq.~(\ref{eq:locking}) is related to the Nambu--Goldstone field by $\R_{\alpha\beta}(x)\sigma_\beta=U^\+(x)\sigma_\alpha U(x)$, which locks the orbital angular momentum density with respect to the spin density.
The latter expression can be expanded as
\begin{align}
s_\alpha(x) = \frac2{\gamma^2}\,\chi_0(x)
\biggl[B_t^\alpha(x) + \frac12\tr[\sigma_\alpha i\d_tU(x)U^\+(x)]
- \R_{\alpha a}(x)\omega_t^a(x) - mg_{ij}(x)v^i(x)v_\alpha^j(x)\biggr] + O(p^3),
\end{align}
where $\chi_0(x)\equiv\chi_0[-\tilde D_t\theta(x)]$ is the local magnetic susceptibility and the temporal component of the spin connection is decomposed as $\omega_t(x)\equiv\omega_t^a(x)\sigma_a$ with
\begin{align}
\omega_t^a(x) = \frac1{4m}e_i^a(x)\varepsilon^{ijk}(x)\d_jA_k(x)
- \frac14\epsilon^{abc}e_b^i(x)\d_te_i^c(x).
\end{align}
We then parametrize the external gauge fields according to $B_t^\alpha(x)\equiv(\gamma/2)H_\alpha(x)$ and $\varepsilon^{ijk}(x)\d_jA_k(x)\equiv2m\Omega^i(x)$, so that $H_\alpha(x)$ is an applied magnetic field and $\Omega^i(x)$ is an angular velocity vector of the rotating frame.
Therefore, we find the Barnett effect where the magnetization is generated by the rotation,
\begin{align}
\gamma s_\alpha(x) = -\frac{\chi_0(x)}{\gamma}\R_{\alpha a}(x)e_i^a(x)\Omega^i(x) + \cdots,
\end{align}
as well as the Einstein--de Haas effect where the orbital angular momentum density is generated by the magnetic field,
\begin{align}\label{eq:density}
\ell_i(x) = -\frac{\chi_0(x)}{\gamma}e_i^a(x)\R_{a\alpha}^T(x)H_\alpha(x) + \cdots.
\end{align}
These gyromagnetic responses are consequences of the spin--orbit locking in the superfluid $^3$He-B~\cite{Salomaa:1987}, which have been long known but are reproduced here in a model-independent and nonperturbative way.
In particular, by taking $-\chi_0/(\gamma\mu_0)\approx4\times10^{-11}$\,Pa\,s/T for $\chi_0\approx10^{-8}$ as an example~\cite{Wheatley:1975}, the magnitude of an orbital angular momentum density induced by a magnetic field of 0.1 tesla is estimated at $\ell\approx4\times10^{-12}$\,Pa\,s.

\subsection{Stress tensor and Hall viscosity}
The intriguing physics of the superfluid $^3$He-B is furthermore reflected in the stress tensor, which is evaluated as
\begin{align}\label{eq:stress}
T^{ij}(x) &= P(x)g^{ij}(x) + v^i(x)j^j(x) + j^i(x)v^j(x) - \rho(x)v^i(x)v^j(x) \notag\\
&\quad - \eta^{ijkl}(x)V_{kl}(x) + O(f_{1-4}) + O(g_{1-4}) + O(p^4),
\end{align}
where $P(x)\equiv\L_\eff(x)$ is the local pressure and $V_{kl}(x)\equiv[\nabla_kv_l(x)+\nabla_lv_k(x)+\d_tg_{kl}(x)]/2$ is the strain rate tensor.
In particular, the viscosity tensor is found to be
\begin{align}\label{eq:viscosity}
\eta^{ijkl}(x) = -\frac{\ell_m(x)}{4}
[\varepsilon^{mik}(x)g^{jl}(x) + \varepsilon^{mjk}(x)g^{il}(x)
+ \varepsilon^{mil}(x)g^{jk}(x) + \varepsilon^{mjl}(x)g^{ik}(x)],
\end{align}
which is the so-called Hall viscosity and is dissipationless because of $\eta^{ijkl}(x)=-\eta^{klij}(x)$~\cite{Avron:1995,Avron:1998,Hoyos:2014b}.
While the Hall viscosity tensor in three dimensions generally bears fifteen independent coefficients~\cite{Avron:1998}, only three of them that constitute a vector under rotation prove to be nonzero and uniquely fixed by the orbital angular momentum density.
It should be remarked that the obtained succinct formula for the Hall viscosity tensor is closely parallel to that of the chiral superfluid in two dimensions~\cite{Read:2009,Read:2011,Hoyos:2014a}.

All the above predictions are expressed in the most general forms in curved space.
They can be readily reduced to the physically relevant ones in flat space by simply setting $e_i^a(x)=\delta_i^a$ and hence $g_{ij}(x)=\delta_{ij}$, $\varepsilon_{ijk}(x)=\epsilon_{ijk}$, and $\nabla_\mu=\d_\mu$.
Furthermore, the ground state is achieved by setting the Nambu--Goldstone fields constant and turning off the spatial components of the external gauge fields so that the mass and spin superfluid velocities vanish.
The remaining external parameters are then the applied magnetic field and the local chemical potential $A_t(x)=\mu(x)$ including a trapping potential.

Finally, in order to investigate an experimental implication resulting from the Hall viscosity, let us consider a uniform state with mass density $\rho_0$ subject to a constant magnetic field, which induces an orbital angular momentum density $\bm\ell$ in a direction fixed spontaneously.%
\footnote{However, when a tiny magnetic anisotropy $\sim-\chi_0^2(\hat\n\cdot\H)^2$ in the energy density resulting from the dipolar coupling is taken into consideration~\cite{Vollhardt:1990}, the rotation axis $\hat\n$ associated with $\R_{\alpha\beta}$ is oriented along $\H$ so that $\bm\ell\parallel\H$ is energetically favored.}
Then, by weakly perturbing the system out of the ground state, a small modulation of the mass density, $\rho'(x)\equiv\rho(x)-\rho_0$, can be produced.
Its propagation as a sound wave in the superfluid $^3$He-B is described by linearized hydrodynamic equations reduced from Eqs.~(\ref{eq:mass}) and (\ref{eq:momentum}) with Eq.~(\ref{eq:stress}):
\begin{align}
\d_t\rho'(x) + \d_ij_i(x) &= 0, \\
\d_tj_i(x) + c_0^2\d_i\rho'(x) &= \frac{\eta^{ijkl}}{\rho_0}\d_j\d_kj_l(x),
\end{align}
where $c_0^2\equiv\d P/\d\rho|_{\rho=\rho_0}$ is the speed of sound.
They are straightforward to solve up to the linear order in $\bm\ell$ by substituting $\rho'(x)=\bar\rho'e^{i\k\cdot\x-i\omega t}$ and $\j(x)=\bar\j e^{i\k\cdot\x-i\omega t}$.
In particular, longitudinal and transverse components of the momentum density are found to be $\bar\j_L=(\omega/k)\bar\rho'\hat\k$ and
\begin{align}\label{eq:transverse}
\bar\j_T = -\frac{i\omega}{2c_0^2\rho_0}\bm\ell\times\bar\j_L,
\end{align}
respectively, with $\omega^2=(c_0k)^2$.
While the dispersion relation is not modified, the sound wave proves to exhibit an elliptical polarization by acquiring the transverse component perpendicular to the orbital angular momentum density~\cite{Nicolis:2011,Barkeshli:2012}.
This is because the transverse force acting on the fluid emerges from the longitudinal velocity modulation through the Hall viscosity.
The transverse to longitudinal amplitude ratio is maximal for $\k\perp\bm\ell$ and is estimated at $\bar j_T/\bar j_L\approx2\times10^{-13}$ by taking $\omega\approx1$\,MHz, $\rho_0\approx100$\,kg/m$^3$, and $c_0\approx300$\,m/s as an example~\cite{Wheatley:1975}.

\section{Summary and conclusion
\label{sec:summary}}
In summary, we established the effective field theory governing the low-energy physics of a superfluid $^3$He-B at zero temperature.
To this end, we first showed that its microscopic action, slightly idealized by neglecting the dipolar coupling and employing the interatomic potential of the Lennard-Jones type, enjoys the $\U(1)_\phi\times\SU(2)_S\times\SO(3)_L$ gauge invariance and the nonrelativistic general coordinate invariance when it is coupled with the external gauge fields and metric.
These symmetry constraints were then exploited to construct the most general effective action written in terms of the Nambu--Goldstone fields up to the next-to-leading order in our power counting scheme.
The obtained effective action proved to include a crucial contribution involving the magnetic susceptibility [Eq.~(\ref{eq:NLO_U0})], from which the following predictions can be extracted:
By applying a small magnetic field, the orbital angular momentum density is generated according to the Einstein--de Haas effect [Eq.~(\ref{eq:density})], which manifests itself uniquely as the circulating mass flow [Eq.~(\ref{eq:current})] and as the dissipationless Hall viscosity [Eq.~(\ref{eq:viscosity})] in the stress tensor.
While the gyromagnetic responses have been long known and can be understood intuitively by the spin--orbit locking in the superfluid $^3$He-B~\cite{Salomaa:1987}, our predictions resulting from the symmetries alone are model-independent, nonperturbative, and hence quantitatively reliable.

In particular, the novel Hall viscosity is the physical quantity that has recently attracted significant interest partly because of its topological nature and universality~\cite{Hoyos:2014b}.
While many theoretical studies have been devoted to the Hall viscosity in two dimensions, little attention has been paid to its closely parallel counterpart in three dimensions.
Here we found that the Hall viscosity of the superfluid $^3$He-B, which is controllable with the applied magnetic field through the Einstein--de Haas effect, polarizes sound wave elliptically by inducing its transverse component in a direction perpendicular to the orbital angular momentum density [Eq.~(\ref{eq:transverse})].
Although the resulting transverse to longitudinal amplitude ratio was estimated to be small, our findings for the superfluid $^3$He-B may promote new efforts toward experimental measurements of the Hall viscosity, which to our knowledge have not been achieved so far in any systems.

\acknowledgments
The authors thank Sho Higashikawa, Masaru Hongo, Takeshi Mizushima, and Ryuji Nomura for valuable discussions.
This work was supported by JSPS KAKENHI Grant Nos.~JP15K17727 and JP15H05855.

\appendix
\section{Next-to-leading-order Lagrangian densities}
Our notation in this paper differs from that in Ref.~\cite{Son:2006} by minus signs in defining $\theta(x)$ and $A_\mu(x)$.
Accordingly, the Lagrangian densities at $O(p^2)$ involving only the phase collective mode adapted for our notation become
\begin{align}
\L_{\theta,1}^{(2)}(x) &= f_1[-\tilde D_t\theta(x)]\,
g^{ij}(x)\d_i\tilde D_t\theta(x)\d_j\tilde D_t\theta(x), \label{eq:NLO_theta1}\\
\L_{\theta,2}^{(2)}(x) &= f_2[-\tilde D_t\theta(x)]
\biggl[\d_t\ln\!\sqrt{g(x)} + \frac{g^{ij}(x)}{m}\nabla_iD_j\theta(x)\biggr]^2, \\
\L_{\theta,3}^{(2)}(x) &= f_3[-\tilde D_t\theta(x)]
\biggl[-\frac{m^2}{4}[2g^{ij}(x)\ddot g_{ij}(x)+\dot g^{ij}(x)\dot g_{ij}(x)]
+ mg^{ij}(x)\nabla_iF_{tj}(x) + \frac14F^{ij}(x)F_{ij}(x) \notag\\
&\qquad - [m\d_t\d_i\ln g(x) - g^{jk}(x)\nabla_j\{m\dot g_{ki}(x)-F_{ki}(x)\}]D^i\theta(x)
+ R_{ij}(x)D^i\theta(x)D^j\theta(x)\biggr], \\
\L_{\theta,4}^{(2)}(x) &= f_4[-\tilde D_t\theta(x)]\,g^{ij}(x)R_{ij}(x), \label{eq:NLO_theta4}
\end{align}
where $R_{ij}(x)\equiv\d_k\Gamma_{ij}^k(x)-\d_i\Gamma_{jk}^k(x)+\Gamma_{ij}^k(x)\Gamma_{kl}^l(x)-\Gamma_{il}^k(x)\Gamma_{jk}^l(x)$, not to be confused with the rotation matrix in the main text, is the Ricci tensor and $f_{1-4}[\,*\,]$ are arbitrary functions~\cite{Son:2006}.
On the other hand, the Lagrangian densities at $O(p^2)$ involving the spin--orbit collective mode are new in this paper and are found to be
\begin{align}
\L_{U,0}^{(2)}(x) &= g_0[-\tilde D_t\theta(x)]\,
\frac12\tr[U^\+(x)i\tilde D_tU(x)U^\+(x)i\tilde D_tU(x)], \\
\L_{U,1}^{(2)}(x) &= g_1[-\tilde D_t\theta(x)]\,
\frac{\varepsilon^{ijk}(x)}{2}\nabla_i\tr[\sigma_j(x)U^\+(x)iD_kU(x)], \label{eq:NLO_U1}\\
\L_{U,2}^{(2)}(x) &= g_2[-\tilde D_t\theta(x)]\,
\frac{\varepsilon^{ijk}(x)}{2i}\tr[\sigma_i(x)U^\+(x)iD_jU(x)U^\+(x)iD_kU(x)], \label{eq:NLO_U2}\\
\L_{U,3}^{(2)}(x) &= g_3[-\tilde D_t\theta(x)]\,
\frac{g^{ij}(x)}{2}\tr[U^\+(x)iD_iU(x)U^\+(x)iD_jU(x)], \\
\L_{U,4}^{(2)}(x) &= g_4[-\tilde D_t\theta(x)]
\biggl[\frac{g^{ij}(x)}{2}\tr[\sigma_i(x)U^\+(x)iD_jU(x)]\biggr]^2. \label{eq:NLO_U4}
\end{align}
Here $g_{0-4}[\,*\,]$ are arbitrary functions but can be identified in the weak coupling limit as $g_0[\mu]=m\kF/(3\pi^2)$, $g_1[\mu]=n/(6m)$, $g_2[\mu]=-n/(10m)$, $g_3[\mu]=-4n/(10m)$, $g_4[\mu]=2n/(10m)$, where $\kF\equiv\sqrt{2m\mu}$ and $n\equiv\kF^3/(3\pi^2)$ are the Fermi momentum and the particle number density, respectively, as functions of the chemical potential $\mu$ without Fermi-liquid corrections~\cite{Vollhardt:1990}.%
\footnote{For the ground state in flat space, Eq.~(\ref{eq:NLO_U1}) leads to the spin current density provided by $j_\alpha^i(\x)=\epsilon^{ijk}\R_{\alpha j}\d_kg_1[\mu(\x)]/2$.
By assuming a uniform system in $x$ and $y$ directions and integrating it from $z=-\infty$ (vacuum) to $z=+\infty$ (bulk), the spin current is found to be $\int\!dz\,j_\alpha^i(\x)=\epsilon^{ijz}\R_{\alpha j}n/(12m)$ in the weak coupling limit, which for $\hat\n=\hat\z$ is consistent with the result of Ref.~\cite{Tsutsumi:2012}.}

We note that other possible candidates, if exist, are not independent of the above Lagrangian densities.
For example,
\begin{align}
\L_U^{(2)}(x) \sim \frac12\tr[g^{ij}(x)\sigma_i(x)U^\+(x)iD_jU(x)]^2
\end{align}
multiplied by an arbitrary function of $\tilde D_t\theta(x)$ complies with the required invariance but can be expressed in terms of Eqs.~(\ref{eq:NLO_U2})--(\ref{eq:NLO_U4}).
Similarly,
\begin{align}
\L_U^{(2)}(x) \sim \frac{\varepsilon^{ijk}(x)}{2}\tr[\sigma_i(x)U^\+(x)G_{jk}(x)U(x)]
\end{align}
multiplied by an arbitrary function of $\tilde D_t\theta(x)$ complies with the required invariance but can be expressed in terms of Eqs.~(\ref{eq:NLO_U1}), (\ref{eq:NLO_U2}), and (\ref{eq:NLO_theta4}) with the help of
\begin{align}
U^\+(x)G_{\mu\nu}(x)U(x) &= iU^\+(x)[D_\mu D_\nu U(x)-D_\nu D_\mu U(x)] + H_{\mu\nu}(x), \\
U^\+(x)D_\mu D_\nu U(x) &= D_\mu[U^\+(x)D_\nu U(x)] + U^\+(x)D_\mu U(x)U^\+(x)D_\nu U(x),
\end{align}
and $\varepsilon^{ijk}(x)\tr[\sigma_i(x)H_{jk}(x)]=-g^{ij}(x)R_{ij}(x)$.
We also note that
\begin{align}
\L_U^{\prime(2)}(x) &\sim \frac{g^{ij}(x)}{2i}\tr[U^\+i\tilde D_tU(x)\sigma_i(x)U^\+(x)iD_jU(x)], \\
\L_U^{\prime(2)}(x) &\sim \frac{g^{ij}(x)}{2}\nabla_i\tr[\sigma_j(x)U^\+i\tilde D_tU(x)], \\
\L_U^{\prime(2)}(x) &\sim \biggl[\frac{m}{2}\dot g_{ij}(x) + \nabla_iD_j\theta(x) + \frac12F_{ij}(x)\biggr]
\frac{g^{ik}(x)g^{jl}(x)}{2}\tr[\sigma_k(x)U^\+(x)iD_lU(x)], \\
\L_U^{\prime(2)}(x) &\sim \biggl[\d_t\ln\!\sqrt{g(x)} + \frac{g^{ij}(x)}{m}\nabla_iD_j\theta(x)\biggr]
\frac{g^{kl}(x)}{2}\tr[\sigma_k(x)U^\+(x)iD_lU(x)] \\
&\!\!\underset{\text{EoM}}{\sim} \biggl[\d_t\tilde D_t\theta(x) + \frac{g^{ij}(x)}{m}D_i\theta(x)\d_j\tilde D_t\theta(x)\biggr]
\frac{g^{kl}(x)}{2}\tr[\sigma_k(x)U^\+(x)iD_lU(x)]
\end{align}
multiplied by arbitrary functions of $\tilde D_t\theta(x)$ are $O(p^2)$ consistent with the gauge invariance and the general coordinate invariance but are incompatible with both the parity and time-reversal invariance.


\begin{thebibliography}{99}

\bibitem{Osheroff:1972a}
D.~D.~Osheroff, R.~C.~Richardson, and D.~M.~Lee,
``Evidence for a new phase of solid He$^3$,''
\href{https://doi.org/10.1103/PhysRevLett.28.885}
{Phys.\ Rev.\ Lett.\ \textbf{28}, 885-888 (1972)}.

\bibitem{Osheroff:1972b}
D.~D.~Osheroff, W.~J.~Gully, R.~C.~Richardson, and D.~M.~Lee,
``New magnetic phenomena in liquid He$^3$ below 3 mK,''
\href{https://doi.org/10.1103/PhysRevLett.29.920}
{Phys.\ Rev.\ Lett.\ \textbf{29}, 920-923 (1972)}.

\bibitem{Vollhardt:1990}
D.~Vollhardt and P.~W\"olfle,
\textit{The Superfluid Phases of Helium 3}
(Taylor \& Francis, London, 1990).

\bibitem{Sigrist:1991}
M.~Sigrist and K.~Ueda,
``Phenomenological theory of unconventional superconductivity,''
\href{https://doi.org/10.1103/RevModPhys.63.239}
{Rev.\ Mod.\ Phys.\ \textbf{63}, 239-311 (1991)}.

\bibitem{Mackenzie:2003}
A.~P.~Mackenzie and Y.~Maeno,
``The superconductivity of Sr$_2$RuO$_4$ and the physics of spin-triplet pairing,''
\href{https://doi.org/10.1103/RevModPhys.75.657}
{Rev.\ Mod.\ Phys.\ \textbf{75}, 657-712 (2003)}.

\bibitem{Lombardo:1999}
U.~Lombardo,
``Superfluidity in nuclear matter,''
\href{https://doi.org/10.1142/9789812817501_0009}
{Int.\ Rev.\ Nucl.\ Phys.\ \textbf{8}, 458-510 (1999)}.

\bibitem{Alford:2008}
M.~G.~Alford, A.~Schmitt, K.~Rajagopal, and T.~Sch\"afer,
``Color superconductivity in dense quark matter,''
\href{https://doi.org/10.1103/RevModPhys.80.1455}
{Rev.\ Mod.\ Phys.\ \textbf{80}, 1455-1515 (2008)}.

\bibitem{Volovik:2003}
G.~E.~Volovik,
\textit{The Universe in a Helium Droplet}
(Oxford University Press, Oxford, 2003).

\bibitem{Mizushima:2015}
T.~Mizushima, Y.~Tsutsumi, M.~Sato, and K.~Machida,
``Symmetry protected topological superfluid $^3$He-B,''
\href{https://doi.org/10.1088/0953-8984/27/11/113203}
{J.\ Phys.:\ Condens.\ Matter \textbf{27}, 113203 (2015)}.

\bibitem{Mizushima:2016}
T.~Mizushima, Y.~Tsutsumi, T.~Kawakami, M.~Sato, M.~Ichioka, and K.~Machida,
``Symmetry-protected topological superfluids and superconductors ---From the basics to $^3$He---,''
\href{https://doi.org/10.7566/JPSJ.85.022001}
{J.\ Phys.\ Soc.\ Jpn.\ \textbf{85}, 022001 (2016)}.

\bibitem{Schnyder:2008}
A.~P.~Schnyder, S.~Ryu, A.~Furusaki, and A.~W.~W.~Ludwig,
``Classification of topological insulators and superconductors in three spatial dimensions,''
\href{https://doi.org/10.1103/PhysRevB.78.195125}
{Phys.\ Rev.\ B \textbf{78}, 195125 (2008)}.

\bibitem{Wilczek:2009}
F.~Wilczek,
``Majorana returns,''
\href{https://doi.org/10.1038/nphys1380}
{Nat.\ Phys.\ \textbf{5}, 614-618 (2009)}.

\bibitem{Okuda:2012}
Y.~Okuda and R.~Nomura,
``Surface Andreev bound states of superfluid $^3$He and Majorana fermions,''
\href{https://doi.org/10.1088/0953-8984/24/34/343201}
{J.\ Phys.:\ Condens.\ Matter \textbf{24}, 343201 (2012)}.

\bibitem{Weinberg:1996}
See, for example, Chap.~19 in
S.~Weinberg,
\textit{The Quantum Theory of Fields}
(Cambridge University Press, Cambridge, 1995).

\bibitem{Son:2006}
D.~T.~Son and M.~Wingate, 
``General coordinate invariance and conformal invariance in nonrelativistic physics: Unitary Fermi gas,''
\href{https://doi.org/10.1016/j.aop.2005.11.001}
{Ann.\ Phys.\ \textbf{321}, 197-224 (2006)}.

\bibitem{Son:2007}
D.~T.~Son,
``Vanishing bulk viscosities and conformal invariance of the unitary Fermi gas,''
\href{https://doi.org/10.1103/PhysRevLett.98.020604}
{Phys.\ Rev.\ Lett.\ \textbf{98}, 020604 (2007)}.

\bibitem{Hoyos:2014a}
C.~Hoyos, S.~Moroz, and D.~T.~Son,
``Effective theory of chiral two-dimensional superfluids,''
\href{https://doi.org/10.1103/PhysRevB.89.174507}
{Phys.\ Rev.\ B \textbf{89}, 174507 (2014)}.

\bibitem{Salomaa:1987}
See Sec.~VII.B in
M.~M.~Salomaa and G.~E.~Volovik,
``Quantized vortices in superfluid $^3$He,''
\href{https://doi.org/10.1103/RevModPhys.59.533}
{Rev.\ Mod.\ Phys.\ \textbf{59}, 533-613 (1987)},
and references therein.

\bibitem{Nakahara:2003}
M.~Nakahara,
\textit{Geometry, Topology and Physics}
(Taylor \& Francis, London, 2003).

\bibitem{Scherer:2003}
See, for example,
S.~Scherer,
``Introduction to chiral perturbation theory,''
\href{https://doi.org/10.1007/0-306-47916-8_2}
{Adv.\ Nucl.\ Phys.\ \textbf{27}, 277-538 (2003)}.

\bibitem{Cooper:2008}
N.~R.~Cooper,
``Rapidly rotating atomic gases,''
\href{https://doi.org/10.1080/00018730802564122}
{Adv.\ Phys.\ \textbf{57}, 539-616 (2008)}.

\bibitem{Hoyos:2012}
C.~Hoyos and D.~T.~Son,
``Hall viscosity and electromagnetic response,''
\href{https://doi.org/10.1103/PhysRevLett.108.066805}
{Phys.\ Rev.\ Lett.\ \textbf{108}, 066805 (2012)}.

\bibitem{Son:2005}
D.~T.~Son,
``Effective Lagrangian and topological interactions in supersolids,''
\href{https://doi.org/10.1103/PhysRevLett.94.175301}
{Phys.\ Rev.\ Lett.\ \textbf{94}, 175301 (2005)}.

\bibitem{Wheatley:1975}
J.~C.~Wheatley,
``Experimental properties of superfluid $^3$He,''
\href{https://doi.org/10.1103/RevModPhys.47.415}
{Rev.\ Mod.\ Phys.\ \textbf{47}, 415-470 (1975)}.

\bibitem{Cross:1975}
M.~C.~Cross,
``A generalized Ginzburg-Landau approach to the superfluidity of helium 3,''
\href{https://doi.org/10.1007/BF01141607}
{J.\ Low Temp.\ Phys.\ \textbf{21}, 525-534 (1975)}.

\bibitem{Ishikawa:1980}
M.~Ishikawa, K.~Miyake, and T.~Usui,
``Intrinsic angular momentum and mass current in superfluid $^3$He-A,''
\href{https://doi.org/10.1143/PTP.63.1083}
{Prog.\ Theor.\ Phys.\ \textbf{63}, 1083-1097 (1980)}.

\bibitem{Mermin:1980}
N.~D.~Mermin and P.~Muzikar,
``Cooper pairs versus Bose condensed molecules: The ground-state current in superfluid $^3$He-A,''
\href{https://doi.org/10.1103/PhysRevB.21.980}
{Phys.\ Rev.\ B \textbf{21}, 980-989 (1980)}.

\bibitem{Combescot:1980}
R.~Combescot and T.~Dombre,
``Nonlinear hydrodynamics in superfluid $^3$He at zero temperature,''
\href{https://doi.org/10.1016/0375-9601(80)90497-1}
{Phys.\ Lett.\ A \textbf{76}, 293-296 (1980)}.

\bibitem{Dombre:1985}
T.~Dombre and R.~Combescot,
``Supercurrent in $^3$He-B,''
\href{https://doi.org/10.1103/PhysRevB.32.1751}
{Phys.\ Rev.\ B \textbf{32}, 1751-1755 (1985)}.

\bibitem{Yip:1986}
S.~Yip,
``The supercurrent and angular momentum in $^3$He-B under a magnetic field,''
\href{https://doi.org/10.1088/0022-3719/19/10/006}
{J.\ Phys.\ C: Solid State Phys.\ \textbf{19}, 1491-1501 (1986)}.

\bibitem{Volovik:1983}
G.~E.~Volovik and V.~P.~Mineev,
``Orbital angular momentum in B phase of $^3$He and its effect on the texture in a rotating vessel,''
\href{http://www.jetpletters.ac.ru/ps/1489/article_22737.shtml}
{JETP Lett.\ \textbf{37}, 127-130 (1983)}.

\bibitem{Volovik:1984}
G.~E.~Volovik and V.~P.~Mineev,
``Gyromagnetism of Cooper pairs in superfluid $^3$He-B,''
\href{http://www.jetp.ac.ru/cgi-bin/index/e/59/5/p972?a=list}
{Sov.\ Phys.\ JETP \textbf{59}, 972-979 (1984)}.

\bibitem{Mineev:1984}
V.~P.~Mineev and G.~E.~Volovik,
``Supercurrent and angular momentum in $^3$He-B induced by magnetic field,''
Proc.\ 17th Int.\ Conf.\ Low Temp.\ Phys.\ \textbf{1}, 39 (1984).

\bibitem{Muzikar:1984}
P.~Muzikar,
``Intrinsic angular momentum in superfluid $^3$He-B,''
Proc.\ 17th Int.\ Conf.\ Low Temp.\ Phys.\ \textbf{1}, 45 (1984).

\bibitem{Mineev:1986}
V.~P.~Mineev,
``Orbital angular momentum in superfluid $^3$He-B in a magnetic field,''
\href{http://www.jetp.ac.ru/cgi-bin/index/e/63/4/p721?a=list}
{Sov.\ Phys.\ JETP \textbf{63}, 721-727 (1986)}.

\bibitem{Hakonen:1987}
P.~J.~Hakonen and K.~K.~Nummila,
``Vortex-free state of $^3$He-B in a rotating cylinder,''
\href{https://doi.org/10.1103/PhysRevLett.59.1006}
{Phys.\ Rev.\ Lett.\ \textbf{59}, 1006-1009 (1987)}.

\bibitem{Korhonen:1990}
J.~S.~Korhonen, A.~D.~Gongadze, Z.~Jan\'u, Y.~Kondo, M.~Krusius, Yu.~M.~Mukharsky, and E.~V.~Thuneberg,
``Order-parameter textures and boundary conditions in rotating vortex-free $^3$He-B,''
\href{https://doi.org/10.1103/PhysRevLett.65.1211}
{Phys.\ Rev.\ Lett.\ \textbf{65}, 1211-1214 (1990)}.

\bibitem{Hakonen:1992}
P.~J.~Hakonen,
``Observations on vortex formation in superfluid $^3$He,''
\href{https://doi.org/10.1016/0921-4526(92)90182-R}
{Physica B \textbf{178}, 83-89, (1992)}.

\bibitem{Avron:1995}
J.~E.~Avron, R.~Seiler, and P.~G.~Zograf,
``Viscosity of quantum Hall fluids,''
\href{https://doi.org/10.1103/PhysRevLett.75.697}
{Phys.\ Rev.\ Lett.\ \textbf{75}, 697-700 (1995)}.

\bibitem{Avron:1998}
J.~E.~Avron,
``Odd viscosity,''
\href{https://doi.org/10.1023/A:1023084404080}
{J.\ Stat.\ Phys.\ \textbf{92}, 543-557 (1998)}.

\bibitem{Hoyos:2014b}
C.~Hoyos,
``Hall viscosity, topological states and effective theories,''
\href{https://doi.org/10.1142/S0217979214300072}
{Int.\ J.\ Mod.\ Phys.\ B \textbf{28}, 1430007 (2014)}.

\bibitem{Read:2009}
N.~Read,
``Non-Abelian adiabatic statistics and Hall viscosity in quantum Hall states and $p_x{+}ip_y$ paired superfluids,''
\href{https://doi.org/10.1103/PhysRevB.79.045308}
{Phys.\ Rev.\ B \textbf{79}, 045308 (2009)}.

\bibitem{Read:2011}
N.~Read and E.~H.~Rezayi,
``Hall viscosity, orbital spin, and geometry: Paired superfluids and quantum Hall systems,''
\href{https://doi.org/10.1103/PhysRevB.84.085316}
{Phys.\ Rev.\ B \textbf{84}, 085316 (2011)}.

\bibitem{Nicolis:2011}
A.~Nicolis and D.~T.~Son,
``Hall viscosity from effective field theory,''
\href{https://arxiv.org/abs/1103.2137}
{arXiv:1103.2137 [hep-th]}.

\bibitem{Barkeshli:2012}
M.~Barkeshli, S.~B.~Chung, and X.-L.~Qi,
``Dissipationless phonon Hall viscosity,''
\href{https://doi.org/10.1103/PhysRevB.85.245107}
{Phys.\ Rev.\ B \textbf{85}, 245107 (2012)}.

\bibitem{Tsutsumi:2012}
Y.~Tsutsumi and K.~Machida,
``Edge current due to Majorana fermions in superfluid $^3$He A- and B-phases,''
\href{https://doi.org/10.1143/JPSJ.81.074607}
{J.\ Phys.\ Soc.\ Jpn.\ \textbf{81}, 074607 (2012)}.

\end{thebibliography}
\end{document}